\documentclass[twocolumn,showpacs,amsmath,amssymb]{revtex4}

\usepackage{latexsym}
\usepackage{graphicx}

\newcommand{\req}[1]{Eq.\,(\ref{#1})}
\newcommand{\beqn}{\begin{equation}}
\newcommand{\eeqn}{\end{equation}}

\begin{document} 
\title{Top anomalous magnetic moment and the two photon decay of Higgs}
\author{Lance Labun and Johann Rafelski}
\affiliation{
Department of Physics, University of Arizona, Tucson, Arizona, 85721 USA \\
TH Division, Physics Department, CERN, CH-1211 Geneva 23, Switzerland}
\date{8 August, 2013}

\begin{abstract}
\vskip -4cm \hspace*{-2.82cm} \noindent CERN-PH-TH/2012-237\vskip 3.4cm 
We compute the dependence of the Higgs to two-photon decay rate $\Gamma_{h\to \gamma \gamma}$ on the top quark gyromagnetic factor $g_t$ in the heavy top limit and evaluate  the expected change for one-loop SM correction to $g_t$. Our results are general and allow consideration of further  modifications of $g_t$ and we predict the resultant $\Gamma_{h\to \gamma \gamma}$.
\end{abstract}

\pacs{12.20.-m, 11.15.Tk, 12.20.Ds, 13.40.-f\\[-0.4cm]}

\maketitle

\begin{figure}[b]
\begin{center}
\begin{picture}(230,100)
\put(0,-5){\includegraphics[width=0.2\textwidth,angle=90]{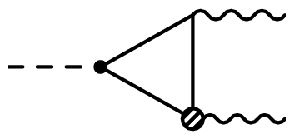}}
\put(37,47){$t$}
\put(26,4){$h$}
\put(10,86){$\gamma$}
\put(50,86){$\gamma$}
\put(75,-5){\includegraphics[width=0.2\textwidth,angle=90]{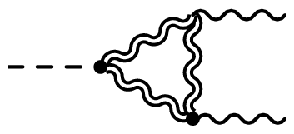}}
\put(86,34){W$^{\pm}$}
\put(150,-5){\includegraphics[width=0.2\textwidth,angle=90]{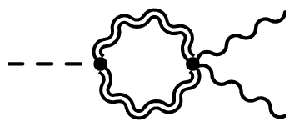}}
\put(183,42){W$^{\pm}$}
\end{picture}
\end{center}
\caption{Diagrams giving dominant contribution to the Higgs $h$ decay to two photons $\gamma$.  In the top loop at left, we denote the top-photon vertex with a shaded circle to signify that we consider a general value of the top gyromagnetic factor.  \label{fig:diags}}
\end{figure}

{\bf Introduction.---}
The recent discovery of a Higgs-like boson~\cite{ATLAS:2012gk,Aad:2013wqa,CMS:2012gu,CMS:ril} provides a new opportunity to study the top quark coupling to photons.  The reason is that the Higgs, being itself uncharged, couples to photons only through loops~\cite{Shifman:1979eb,Ellis:1975ap,Marciano:2011gm}.  The dominant contributions due to the W boson and top quark  are illustrated in Fig.~\ref{fig:diags}.  In determination of the relevant contributions from fermions, there is a competition between the fermion mass, which reduces loop strength, and the fermion-Higgs coupling, which is proportional to mass.  Therefore, the top quark contributes a significant fraction of the Higgs-two-photon amplitude.

The possible deviation of the decay rate $\Gamma_{h\to\gamma\gamma}$ from the standard model (SM) prediction has generated a lot of interest in finding a mechanism for a deviation in ${h\to\gamma\gamma}$ that minimally affects other SM results. We contribute to this objective by  evaluating the   Higgs decay rate $\Gamma_{h\to\gamma\gamma}$ as a function of  the top quark gyromagnetic ratio $g_t$, related in conventional way to the particle magnetic moment
\beqn\label{mutdefn}
\mu_t=\frac{g_t}{2}\frac{Qe}{2m_t}, \quad Qe=+(2/3)e
\eeqn 
The magnetic moment is neither fixed nor protected by a conservation law: the Dirac gyromagnetic ratio $g_{\rm D}=2$ for bare point-like fermions  is well-known to be modified by quantum corrections.  The coupling of top to Higgs, $y_{th}=m_t\sqrt{2}/v\simeq 0.99$, [in which the Higgs vacuum expectation value $v=246.2\,\mathrm{GeV}\simeq (G_F\sqrt{2})^{-1/2}$, $G_F$ the Fermi constant  and $m_t=173.4$ GeV] could herald top structure capable of greatly altering the value of  $g_t$.

The lowest order SM calculation of $g_t$ yields at one loop three contributions: a) the standard QED result, b) a similar QCD result~\cite{Bernreuther:2004ih,Bernreuther:2005gq}, and c) electroweak contribution:
\begin{subequations}\label{1loopg}
\begin{align}
(g_t-2)^{(1L)}_{\rm QED} &= \alpha_{\rm QED}/\pi =2.5 \:10^{-3}  \\
(g_t-2)^{(1L)}_{\rm QCD} &= \frac{N_c^2-1}{2N_c}\frac{\alpha_s}{\pi} =25 \:10^{-3} \\
(g_t-2)^{(1L)}_{\rm EW}  &= 7.5\:10^{-3}.
\end{align}
\end{subequations}
To obtain the EW numerical value  we note that all vertex corrections  have the same  loop integral. Therefore, we can use Eq.\,(3) of~\cite{Martinez:2007qf} in combination with their Eq.\,(1) which identifies their $\Delta\kappa\to g_t-2$. Both the virtual gluon and Higgs exchange corrections to $g_t$ can be relatively large   considering the coupling strengths $\alpha_s(m_t)\simeq 0.11$ and $y_{th}^2/4\pi=0.08$.

Furthermore, composite particles can have gyromagnetic ratios unrelated to $g_{\rm D}$. The sensitivity of the magnetic moment  to compositeness is well known~\cite{Brodsky:1980zm}. Determining $\mu_t$ will provide important information about the structure of the top, whether it is point-like or composite and whether or not it couples with beyond the standard model (BSM) particles~\cite{Bernreuther:2008ju,Kamenik:2011dk,Bouzas:2012av}. 

Many efforts have focused on relating the anomalous moments to top quark production and decay~\cite{Atwood:1991ka,Atwood:1994vm,Haberl:1995ek,Zhang:2010dr,Larkoski:2010am}, and the radiative decay of the bottom to strange quark $b\to s\gamma$ has also long been studied in search of indirect constraints~\cite{Hewett:1993em,Martinez:1996cy} on $g_t$.  The experimental data continues to be analyzed to place constraints on the dipole moments~\cite{Kamenik:2011dk,Bouzas:2012av} and compositeness~\cite{Fabbrichesi:2013bca}.  Recently new opportunities provided by Higgs production and decay~\cite{Choudhury:2012np,Degrande:2012gr,Elias-Miro:2013gya,Grojean:2013kd} have been recognized.

The Higgs-two-photon decay amplitude receives a contribution from the top given by the low-energy theorems of~\cite{Shifman:1979eb,Ellis:1975ap}, evaluated employing  the effective action
\beqn
\label{lowELeff}
{\cal L}_{h\gamma\gamma}=
  -b_0\frac{\alpha}{2\pi}\frac{h}{v}\frac{1}{4}F^{\mu\nu}F_{\mu\nu}
\eeqn
to leading order in the electromagnetic coupling constant $\alpha$ and in the ratio $m_h/2m_t\simeq 0.36$ for $m_h\simeq 125.5~{\rm GeV}$.  Here $F^{\mu\nu}$ is the electromagnetic field strength tensor, and $b_0$ is the one-loop beta-function coefficient in the expansion of the renormalization group: ${-d\alpha}/{d\ln \lambda}\equiv -\beta(\alpha)=(b_0/2\pi)\alpha^2+ (b_1/4\pi^2) \alpha^3\ldots$, where the scale $\lambda$ is fixed by measuring the coupling $\alpha(\lambda)$.   In the form \req{lowELeff} the combination $\beta(\alpha)F^{\mu\nu}F_{\mu\nu}/4\alpha$ is renormalization group invariant to one loop, and hence no logarithmic corrections arise~\cite{Elias-Miro:2013gya}.

In fact, the effective Lagrangian \req{lowELeff} gives the amplitude to within $3\%$ of the loop calculation~\cite{Marciano:2011gm}.  Therefore, by obtaining $b_0(g_t)$, which further displays the effect of the top quark magnetic moment on the polarization of the photon, we also obtain the $h\to\gamma\gamma$ amplitude to required precision.  Corrections in powers of $1/m_t$~\cite{Bernreuther:1983} can be computed separately, as they are for the Higgs-gluon effective coupling, see e.g.~\cite{Harlander:2009mq}.  

\vspace{0.2cm}\noindent{\bf Describing top anomalous magnetic moment.---}
The difficulty of handling an effective non-Dirac magnetic moment is well known.  We are aware of three possible approaches: 
\begin{enumerate}
\item   The perturbative evaluation of $g_t$ can be continued, summing contributions in the theory based at $g_t=2$, adding  higher order diagrams to the top loop seen in Fig.\ref{fig:diags}. It is known that perturbative expansion is slowly convergent~\cite{Bernreuther:2004ih}. Moreover, this approach excludes the possibility of testing for composite structure by considering arbitrary $g_t$.
\item   One can complement the Dirac action for the top quark  with an effective operator of mass dimension 5 (or 6, if before electroweak symmetry breaking). This procedure  introduces by dimensional counting a logarithmic divergence.  The logarithm can be connected with the running of the dimensionful coupling constant from the BSM scale  to the Higgs' scale, and the effect of this running is being discussed by several groups~\cite{Degrande:2012gr,Grojean:2013kd,Elias-Miro:2013gya}. 
\item  Here we consider the top quark as effectively pointlike at the scale of study,  $m_h$. Therefore we employ a description that is independent of additional scales. This is achieved by introducing the second-order theory of fermions, in essence  `squaring'  the Dirac operator  following a method first proposed by Schwinger~\cite{Schwinger:1951nm}.  The resulting doubling of Fermi degrees of freedom can be dealt with by including where  appropriate  a factor 1/2. However,  further refinement of the method is necessary for $g_t>2$~\cite{Rafelski:2012ui}.
\end{enumerate}

The action we consider for the top coupled to the electromagnetic field is
\beqn\label{Ltop}
{\cal L}_t= \bar t \left(\Pi^2-m_t^2-\frac{g_t}{2}\frac{e\sigma_{\mu\nu}F^{\mu\nu}}{2}\right)t
\eeqn
with $\Pi^{\mu}=i\partial^{\mu}-eA^{\mu}$ the Hermitian momentum operator. The fermion fields have mass dimension 1.  Hence the magnetic moment interaction seen in \req{Ltop} is  dimension-4 operator. Because the arbitrary magnetic moment is renormalizable in this framework, we obtain a finite well-determined result for the low-energy effective coupling $b_0$ for arbitrary $g_t$, and hence for the $h\to\gamma\gamma$ amplitude. 

Equation (\ref{Ltop}) has many early mentions in literature discussed in~\cite{AngelesMartinez:2011nt}.  At the point $g_t=2$, the 2nd-order theory produces exactly  equivalent results to the 1st order Dirac action for quantum electrodynamics (QED) (results agree up to the factor 1/2 mentioned above needed to correct the number of degrees of freedom e.g. in fermion loops), and to one loop in perturbation theory it has been shown to be renormalizable for all $g_t$~\cite{VaqueraAraujo:2012qa}. A second order formulation of the entire Standard Model has recently been discussed~\cite{Espin:2013wia}. 

Using a second order theory of the top, there is a further diagram in Fig.\,\ref{fig:diags} corresponding to two-top-two-photon coupling analogous to the $W$-loop in (c).  This contribution is included when deriving the low-energy theorem and evaluating the top loop amplitude but is not shown explicitly in Fig.\,\ref{fig:diags}.

\vspace{0.2cm}\noindent{\bf $g_t$-dependent effective coupling.---}
The one-loop $\beta$-function coefficient $b_0$ has been obtained in the second order formulation for a fermion with arbitrary magnetic moment using two independent methods: a) perturbative computation~\cite{AngelesMartinez:2011nt} and b) external field  method~\cite{Rafelski:2012ui}. The perturbative result for the vacuum polarization top-quark loop is (Eq.\,(3.24) of Ref.~\cite{VaqueraAraujo:2012qa})
\beqn\label{betafuncpert}
b_0(g_t) = -\frac{4}{3}N_cQ^2\!\left(\frac{3}{8}g_t^2-\frac{1}{2}\right).
\eeqn
The  $Q^2=4/9$ arises from the charge of the top, and $N_c=3$ for the color trace. The $-4/3$ factor separated in front is the well-known value of $b_0$ for a fermion with unit charge at  $g_t=2$.  

The low-energy theorem \req{lowELeff} was originally  derived from the form of the effective potential obtained in the external field method~\cite{Shifman:1979eb}. When using the external-field method an in-depth study of $|g_t|>2$ is necessary following the observation that Schwinger proper-time evaluation of the effective action~\cite{Schwinger:1951nm} when adapted to arbitrary value of $g_t$~\cite{Labun:2012jf} diverges for $|g_t|>2$. This suggests that the convergence radius of the perturbative  $b_0(g_t)$, \req{betafuncpert}, is   $|g_t|\le 2$.

A convergent result for the effective potential is achieved when the doubled degrees of freedom are properly separated into two half-Hilbert spaces for $|g_t|>2$~\cite{Rafelski:2012ui}.  The separation accomplishes stability and Lorentz invariance of the top quark vacuum, and the procedure is repeated at  $g_t^{(N)}=2+4N$ for each integer $N$. At these values $g_t^{(N)}$ there is a countably infinite number of level crossings occurring between states belonging to the two half-Hilbert spaces.  As a result of this procedure, $b_0(g_t)$ is found  to be a periodic function, repeating the fundamental domain $-2\leq g_t\leq 2$ for $|g_t|>2$, see  in Eq.\,(9) of Ref.~\cite{Rafelski:2012ui}, and for further discussion see  Sec. II of~\cite{Labun:2012ra}:   
\begin{align}\label{betafunc}
b_0(g_t) &=-\frac{4}{3}N_cQ^2\!\left(\frac{3}{8}(g_t\!-\!4N)^2-\frac{1}{2}\right)
\\ &-2\leq g_t-4N\leq 2,~~N\in \mathbb{Z} \notag
\end{align}
The perturbative result for $b_0(g_t)$ \req{betafuncpert} corresponds to fixing $N=0$ and ignoring the periodicity.

An important and unexpected observation about $b_0(g_t)$, first made in Ref~\cite{Rafelski:2012ui}, is that it changes sign at $g_t-4N=\pm 2/\sqrt{3}$ and hence is positive for $|g_t-4N|<2/\sqrt{3}$.  For the perturbative evaluation of $b_0(g_t)$, this sign change occurs twice, when $N=0$.  The consideration of diamagnetic (originating in $\Pi^2$) and paramagnetic (originating in $\sigma_{\mu\nu}F^{\mu\nu}$) contributions in \req{betafunc} reveals  that this effect is due to the decrease in strength of the paramagnetic terms~\cite{Nink:2012vd}. As $g_t^2$ diminished, the paramagnetic component in $b_0(g_t)$ diminishes,  and the  `asymptotic safety' disappears, see  Eq.\,(10) of~\cite{Reuter:MG13}.

\vspace{0.2cm}\noindent{\bf Higgs to two photon rate.---} 
The top quark loop contribution to the  $h\to \gamma \gamma$  decay is derived from \req{lowELeff} by inserting photon momentum and polarization vectors for each $F^{\mu\nu}$ in \req{lowELeff}~\cite{Shifman:1979eb} (see also \cite{Ellis:1975ap}),
\beqn\label{tcontrib}
A_t(h\to \gamma\gamma)\simeq \frac{1}{v}\frac{\alpha\,b_0}{4\pi}(k^{\kappa}_1\epsilon^{\lambda}_1-k^{\lambda}_1\epsilon^{\kappa}_1)(k^{\kappa}_2\epsilon^{\lambda}_2-k^{\lambda}_2\epsilon^{\kappa}_2)
\eeqn
$k^{\kappa}_{1,2}$ and $\epsilon^{\lambda}_{1,2}$ are the 4-momenta and polarization vector of the photons.  Evaluating the amplitude from~\req{tcontrib} means an error of  a few percent relative to the result from the loop amplitude~\cite{Shifman:1979eb,Marciano:2011gm} for the value  $(m_h/2m_t)^2\simeq 0.13$.

The dependence of the total $h\to \gamma\gamma$ decay amplitude on the top magnetic moment is found by combining the $g$-dependent top-loop contribution with the contribution from the W-boson loop.  Inserting the $g$-dependent $\beta$ function \req{betafunc} in \req{tcontrib}, the total amplitude for Higgs decay into two photons is 
\beqn\label{h2gamma}
A_{\rm tot}(h\to \gamma\gamma)\simeq A_t(h\to \gamma\gamma)+A_W(h\to \gamma\gamma)
\eeqn
with the W loop contributing~\cite{Shifman:1979eb,Marciano:2011gm} 
\begin{align}\label{Wamplitude}
&A_W(h\to \gamma\gamma)= f_W\frac{1}{v}\frac{\alpha}{4\pi}(k^{\kappa}_1\epsilon^{\lambda}_1\!-\!k^{\lambda}_1\epsilon^{\kappa}_1)(k^{\kappa}_2\epsilon^{\lambda}_2\!-\!k^{\lambda}_2\epsilon^{\kappa}_2) \\  &f_W(x)=3x(2-x)\left(\arcsin(x^{-1/2})\right)^{\!2}\!+3x+2,~~x=\frac{4m_W^2}{m_h^2} \notag
\end{align}
for which we have stated only the relevant case $4m_W^2/m_h^2\equiv x>1$ from~\cite{Shifman:1979eb}, and we use  $x=1.641$ corresponding to $m_h=125.5$ GeV.  A similar function $f_q(x)$ (see Eq.\,(21) in~\cite{Shifman:1979eb}) derived from the fermion loop generalizes the low energy \req{tcontrib}  and with $x_t=4m_t^2/m_h^2\to 7.64$ gives a numerical value differing by less than 4\% from the low energy ($x\to\infty$) result, as mentioned above.

To compare with the measured decay rate, we evaluate the decay width (Eq.\,(3) of~\cite{Shifman:1979eb} and Eq.\,(7) of~\cite{Marciano:2011gm}
\beqn\label{decayrate}
\Gamma_{h\to \gamma\gamma}\simeq \left| f_W(\frac{4m_W^2}{m_h^2})+b_0(g_t) \right|^2\left(\frac{\alpha}{4\pi}\right)^{\!2}\frac{m_h^3}{16\pi v^2}, 
\eeqn
with $f_W(x)$ given in \req{Wamplitude} and $b_0(g)$ from \req{betafunc}.  Equation (\ref{decayrate})  provides  for $g_t\to 2$ the Higgs to two photon decay rate within a few percent of the width stated in the 2011 updated Higgs Cross Section Working Group tables  (which include NLO QCD and electroweak corrections), after accounting for partial  decay widths   $h\to ZZ$, $h\to WW$ and to $4$-fermions, see Eq.\,(1) of~\cite{Denner:2011mq}.

\begin{figure}[t]
\centering
\includegraphics[width=0.48\textwidth]{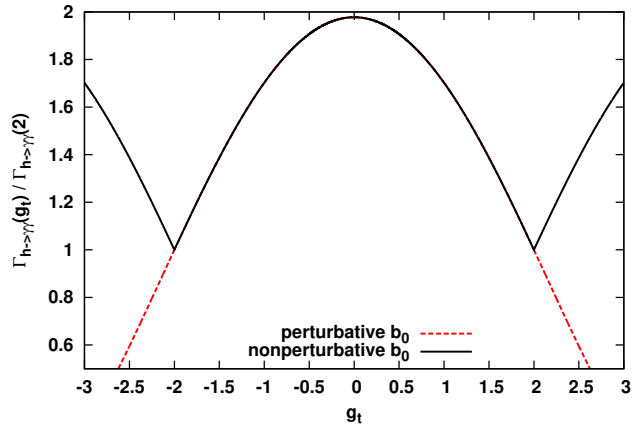}
\caption{Higgs to two photons $h\to \gamma\gamma$ decay rate \req{decayrate} normalized to its value at $g_t=2$ for $x=4m_W^2/m_h^2=1.64, m_h=125.5$  GeV.  The dashed (red) line corresponds to the perturbative evaluation~\cite{AngelesMartinez:2011nt} of $b_0(g_t)$, \req{betafuncpert} and the solid (black) line corresponds to the external field method~\cite{Rafelski:2012ui}  result \req{betafunc}.
\label{fig:amplitude}}
\end{figure}

Figure~\ref{fig:amplitude} shows the $g_t$ dependence of the total  $h\to \gamma\gamma$ decay rate, normalized to its value at $g_t=g_D=2$ for both non-perturbative and perturbative evaluation.  Considering the periodicity of $b_0(g_t)$  the rate $\Gamma_{h\to \gamma\gamma}$ is always enhanced in non-perturbative approach, as compared to $g=g_{\rm D}$ (and its periodic recurrences i.e   $g_t=\pm 2,\pm 6, \pm 10$ etc). However, the prediction using the perturbative computation of $b_0(g_t)$  implies  for $|g_t|>2$ a suppression of $h\to\gamma\gamma$.
 
The line of reasoning here applied to $h\to\gamma\gamma$ also applies to Higgs decays into 2 gluons, $h\to GG$, because the $hGG$ interaction is similarly an effective interaction involving a top quark loop and to a smaller degree a bottom quark loop.  In contrast to the $h\to \gamma \gamma$  case, in the $GG$ decay channel the top quark loop gives the dominant contribution.  The $h\to GG$ dependence on chromomagnetic moment $g^c_t$ and related experimental opportunities are discussed in~\cite{Labun:2012ra}. Although similar SM and beyond Standard Model (BSM) physics can contribute to both $g^c_t$ and $g_t$, it is important to remember that there is no immediate relation between the specific values of $g^c_t$ and~$g_t$.

\vspace{0.2cm}\noindent{\bf Comparison of $g_t$  with experiment.---}
Experimental input on the magnitude of $g_t$ is from here considered Higgs decay,  the radiative $b\to s\gamma$ decay, and   top  production. 

{\it Higgs$\to\gamma\gamma$ decay:}
This study was prompted by reports of possible enhancement of Higgs candidate $\to\gamma\gamma$ decay: ATLAS recently released a report of a $\sim 55\%$ enhancement over the SM prediction in the $h\to\gamma\gamma$ rate~\cite{Aad:2013wqa}.   The combined Tevatron results~\cite{Aaltonen:2013kxa} also find rate enhancement at  $1\sigma$ level.   CMS initially reported an enhancement~\cite{CMS:2012gu}, but in a later  update~\cite{CMS:ril} concludes there is a possible slight suppression,  agreeing at $1\sigma$ with the SM prediction.

Based on 1-loop result \req{1loopg} for $g_t$ and evaluating the amplitude \req{h2gamma}, we find not more that a 1.0\% top decay amplitude modification and hence a 2.0\% modification of the decay rate \req{decayrate}.  This  suggests that  the SM perturbative value of $g_t-2$ predicts too small a modification (a few percent) of decay rate to explain a potential enhancement as reported by ATLAS. However, there are further contributions of higher order QCD  and virtual Higgs which have not been considered in \req{1loopg} as well as BSM effects which may be visible in $g_t$.

{\it The radiative $b\to s\gamma$ decay:} The earlier work~\cite{Hewett:1993em,Martinez:1996cy} was recently updated, see Eq.\,(9) in~\cite{Kamenik:2011dk}
\begin{align} -1.83<\Delta\mu_tm_t<0.53, 
\qquad \frac{\Delta\mu_t}{2}=\frac{g_t-2}{2}\frac{\frac{2}{3}e}{2m_t}
\end{align}
In above   we have not considered the electric dipole moment term, which is constrained to be $\sim 10^6$ times smaller than $\mu_t$.   We consider the numbers  in Eq.\,(9) of~\cite{Kamenik:2011dk} dimensionless and translating into a bound on $g_t$, 
\beqn
-3.49 <g_t<3.59 \label{gexplimits}
\eeqn
This is thus not constraining in our consideration, as it is consistent  with the full range of $g_t$ factors suggested by the enhancement of the rate shown in figure~\ref{fig:amplitude}.

{\it top  production:} Limits on $g_t$ should arise from the study of top production. In hadron colliders, top production is predominantly via strong interactions, so the LHC is more sensitive to the top chromomagnetic moment, i.e. the corresponding $g^c_t$ ($c$ for chromodynamic factor)~\cite{Hioki:2009hm}, and $g^c_t$ could perhaps be connected using Schwinger-Dyson equations  to $g_t$.  Only a future direct study of the top production in $e^+e^-$ collisions appears to offer another   sensitive measure  of the QED top anomalous magnetic dipole moment~\cite{Atwood:1991ka,Devetak:2010na}.

\vspace{0.2cm}\noindent{\bf Summary and conclusions.---}
We have demonstrated the decay rate $h\to \gamma\gamma$ depends significantly on the top quark gyromagnetic ratio $g_t$, presenting two results:\\ 
1)  The perturbative result follows from a) the separation of $b_0$ into its paramagnetic and diamagnetic components (see section 12.3.3 of~\cite{Huang:1982ik}), and b) can be obtained in explicit Feynman diagram evaluation~\cite{VaqueraAraujo:2012qa}.  Using this perturbative form we presented the resultant perturbative $h\to \gamma\gamma$ decay rate. \\
2) However,  one must expect that the perturbative evaluation has a finite radius of convergence because the strength of paramagnetic interaction increases as $g_t^2$. In the external-field method we identified a divergence of effective action~\cite{Labun:2012jf} for  $|g_t|> 2$.  We believe that appearance of this divergence sets the radius of convergence of the perturbative method at $|g_t|\le 2$. The proposed solution yielding a finite effective action~\cite{Rafelski:2012ui} leads to  periodic behavior of $h\to  \gamma\gamma$ decay rate  as a function of $g_t$ valid for all $g_t$.

This non-perturbative result \req{betafunc} implies an enhancement of $h\to\gamma\gamma$ for any $g\ne 2$, see Figure~\ref{fig:amplitude}. The perturbative evaluation in the domain $|g_t|>2$ produces  a suppression of the $\Gamma_{h\to\gamma\gamma}$ in  the SM expected $g_t-2>0$ domain. Both  results are finite since we are using the `squared' Dirac operator in computations. In this second-order theory of the top quark, the magnetic moment interaction is renormalizable for any  $g_t$, and no discussion of the scale-dependence or running enters~\cite{Degrande:2012gr,Grojean:2013kd,Elias-Miro:2013gya}.  

Irrespective whether the perturbative or the external-field result applies, the  rate $\Gamma_{h\to\gamma\gamma}$ offers the greatest sensitivity to top quark magnetic properties. When complemented with a further independent measurements of $g_t$, a study of  $\Gamma_{h\to \gamma \gamma}$ could experimentally settle the theoretical question about the behavior of $\beta(g_t)$ for large $g_t$. In view of the  $g_t$-dependence of the rate $\Gamma_{h\to\gamma\gamma}$ presented in Fig.\ref{fig:amplitude}, the Higgs-two-photon decay $h\to\gamma\gamma$ may be offering the most precise presently available information on $g_t$, which will reflect composite or other top quark structure and thus could be a sensitive probe of BSM physics.  

A today visible value $g_t\ne 2$  implies appearance of BSM top quark structure and the $h\to \gamma\gamma$ decay turns out to be an excellent probe of this. The question arises how $g_t\ne 2$ affects all the other observables; however, it is beyond the scope of this work to study the slight tension that exists between the directly measured Higgs mass value, $m_h=125.5$\,GeV and the  fit to precision electro-weak experimental data that predicted  a considerably smaller Higgs mass~\cite{Djouadi:2005gi}  $m_h\simeq 100$\,GeV.

\acknowledgments

We thank the  TH-division of the CERN Physics Department for hospitality while  much of this work was done in 2012.  This work was supported by a grant from the  US Department of Energy, DE-FG02-04ER41318.


\end{document}